\documentstyle[12pt,a4,epsf,graphicx]{article}

\newcommand{\be}{\begin{equation}}
\newcommand{\ee}{\end{equation}}
\newcommand{\ba}{\begin{eqnarray}}
\newcommand{\ea}{\end{eqnarray}}
\newcommand{\dis}{\displaystyle}

\newcommand{\nn}{\nonumber}

\begin{document}
\begin{titlepage}
\begin{flushright}
CAFPE-41/04\\
LU TP 05-3\\
UG-FT-171/04\\
\end{flushright}
\vspace{2cm}
\begin{center}

{\large\bf The $B_K$ Kaon Parameter in the $1/N_c$ Expansion
\footnote{
Invited talk given by J.P.
at `` Large $N_c$ QCD Workshop'', 5-9 July 2004, Trento, Italy.}}\\
\vfill
{\bf  Joaquim Prades$^{a)}$,
Johan Bijnens$^{b)}$, 
 and Elvira G\'amiz$^{c)}$ 
}\\[0.5cm]
$^{a)}$ Centro Andaluz de F\'{\i}sica de las Part\'{\i}culas
Elementales (CAFPE) and Departamento de
 F\'{\i}sica Te\'orica y del Cosmos, Universidad de Granada \\
Campus de Fuente Nueva, E-18002 Granada, Spain.\\[0.5cm]

$^{b)}$  Department of Theoretical Physics, Lund University\\
S\"olvegatan 14A, S-22362 Lund, Sweden.\\[0.5cm]

$^{c)}$ Department of Physics \& Astronomy,
University of Glasgow\\ Glasgow G12 8QQ, United Kingdom.\\[0.5cm]

\end{center}
\vfill
\begin{abstract}
\noindent
We present work going on calculating
the kaon $B_K$ parameter in the $1/N_c$ expansion.
The goal of this work is to analyze analytically 
in the presence of  chiral corrections  this 
phenomenologically very important parameter.
 We present the method used and  preliminary results
for the chiral limit value for which we get 
$\hat B_K^\chi=0.29 \pm 0.15$.
We also give some analytical indications of why 
the large $N_c$ prediction of $\hat B_K= 3/4$
may have small $1/N_c$ corrections in the real case.
\end{abstract}
\vfill
January 2005
\end{titlepage}

\section{Introduction}

  Indirect Kaon CP-violation  in the Standard Model
(SM) is  proportional to the matrix element
\ba
\label{matrix}
\langle \overline K^0 | K^0 \rangle
&=& -i C_{\Delta S=2} \, C(\nu) \, \langle \overline{K^0}|
\int {\rm d}^4 y \, Q_{\Delta S=2} (y) | K^0 \rangle \nn \\
&\equiv&-i C_{\Delta S=2} \, \frac{16}{3} \, \hat B_K \, f_K^2 m_K^2
\ea
with
\ba
Q_{\Delta S=2}(x) \equiv 4 L^\mu(x) L_\mu(x) \, ;
\hspace*{0.5cm} 2 L_\mu(x) \equiv
\left[\overline s \gamma_\mu (1-\gamma_5) d\right](x) \, .
\ea
and $C(\nu)$ is a Wilson coefficient
which is known in perturbative QCD at next-to-leading (NLO) order
in $a\equiv \alpha_S /\pi$  in two schemes \cite{NLO}, namely, the 
't Hooft-Veltman (HV) scheme ($\overline{\rm MS}$ subtraction
and non-anti-commuting $\gamma_5$ in $D\neq 4$)
and in the Naive Dimensional Regularization (NDR) scheme
($\overline{\rm MS}$ subtraction
and anti-commuting $\gamma_5$ in $D\neq 4$). 
The coefficient $C_{\Delta S=2}$ collects
well known functions of the integrated out heavy particle
masses and Cabibbo-Kobayashi-Maskawa matrix elements. For comprehensive
reviews where these factors and 
complete details can be found see \cite{rev}.
The Wilson coefficient $C(\nu)$ is 
\be
C(\nu)= \left( 1 + a(\nu) \left[\frac{\gamma_2}{\beta_1}-
\frac{\beta_2 \gamma_1}{\beta_1^2} \right]\right)\, 
[ \alpha_s(\nu) ]^{\gamma_1/\beta_1}
\ee
where $\gamma_1$ is the one-loop $\Delta S=2$ anomalous dimension
\be
\gamma_1 = \frac{3}{2}\left(1-\frac{1}{N_c}\right) \, ; 
\ee
$\gamma_2$ is the two-loop $\Delta S=2$ anomalous dimension
\cite{NLO}
\ba
\gamma_2^{\rm NDR}&=&
-\frac{1}{32}\left(1-\frac{1}{N_c}\right)
\left[17 + \frac{4}{3} (3-n_f)
+ \frac{57}{N_c}\left(\frac{N_c^2}{9}-1\right)\right]
\, , \nn \\ 
\gamma_2^{\rm HV}&=&
\gamma_2^{\rm NDR}-\frac{1}{2}\left(1-\frac{1}{N_c}\right) \beta_1  
\ea
and  $\beta_1$ and $\beta_2$ are the first two coefficients
of the QCD beta function.

The so-called $\hat B_K$ kaon parameter defined 
in (\ref{matrix})  is an important
input for the unitarity triangle analysis 
 and its calculation has been addressed  many times in the past.
 There have been four main  techniques used to calculate
the $\hat B_K$ parameter: QCD-Hadronic Duality \cite{PR85,PRA91},
three-point function QCD Sum Rules \cite{threepoint},
lattice QCD and the $1/N_c$ ($N_c =$ number of colors)
expansion. For a recent review on the unitarity
triangle where  the relevant references for the inputs
  can be found see \cite{BAT03}.
For  recent advances  using  lattice QCD see \cite{lattice04a,lattice04b}.

Here, we would like to present work going on
determining the $\hat B_K$ parameter at NLO order in the
 $1/N_c$ expansion. That the $1/N_c$ expansion  would be useful in this 
regard was first suggested by Bardeen, Buras and G\'erard
\cite{BBG} and reviewed by Bardeen in \cite{BAR89,BAR99}. There
one can find most of the references to previous work
and applications of this non-perturbative technique.

 Recent related work to the one we will discuss in the following
sections can be found in \cite{BP95,BP00} where a NLO in  
$1/N_c$ calculation of $\hat B_K$ within  and outside
the chiral limit is presented. There, the relevant spectral function 
is calculated using the ENJL  model \cite{ENJL} 
at intermediate energies  while at very low energies  and at very large
energies the chiral perturbation theory (CHPT) 
and operator product expansion OPE) results,respectively,  are used.
 Another calculation of  $\hat B_K$  in the chiral limit
at NLO in the $1/N_c$  is in \cite{PR00}. There, 
the relevant spectral function is saturated by the pion
pole  and the  first rho meson resonance --minimal hadronic 
approximation (MHA). In \cite{CP03}, the same technique as 
in \cite{PR00} was used  but including  also 
the effects of dimension eight operators in the OPE
of the $\Delta S=2$ Green's function and adding  the first
scalar meson resonance to the relevant spectral function.

\section{Technique}
\label{Technique}

All the details on the $X$-boson method were given in \cite{BP00}.
In particular, in that reference it was shown how short-distance
scale and scheme dependences can be taken into account analytically 
in  the $1/N_c$ expansion. Here, we just sketch the procedure
introducing the notation. We want to calculate the two-point function
\cite{BP95,BP00}
\ba
\label{twopoint}
{\bf \Pi}_{\Delta S=2}(q^2)
= i {\dis \int} {\rm d}^4 \, e^{i q \cdot x}\, 
\langle 0 | T \left(P_{\overline{K}^0}^\dagger(0)
P_{K^0}(x) \, e^{i {\bf \Gamma}_{\rm LD}}
\right) | 0 \rangle 
\ea
in the presence of the long-distance
$\Delta S=2$ effective action of the Standard Model 
${\bf \Gamma}_{\Delta S=2}$.
 After reducing the kaon two-quark densities the two-point function
(\ref{twopoint}) provides the matrix element in (\ref{matrix}).

The effective action ${\bf \Gamma}_{\Delta S=2}$ reproduces the
physics of the SM at low energies by the exchange of a colorless
heavy $\Delta S=2$ $X$-boson. To obtain it \cite{BP00}, we make
a short-distance matching analytically between
\be
{\bf \Gamma}_{\Delta S=2}
\equiv - C_{\Delta S=2} \, C(\nu)  \int {\rm d}^4 y \, Q_{\Delta S=2}(y)
+ {\rm h.c.} 
\ee
and --in our approach--
\be
{\bf \Gamma}_{\rm LD}
\equiv 2 \, g_{\Delta S=2}(\mu_C,\cdots)\, 
  \int {\rm d}^4 y \, X^\mu(y) \, L_\mu(y) \, + {\rm h.c.} 
\ee
Where, we have chosen to regulate the long-distance
effective action in four dimensions with a cut-off
$\mu_C$. These choices are perfectly compatible with keeping the
short-distance scale and scheme independence
analytically exact and we believe that  at low energies,
where the relevant degrees of freedom are not quarks and gluons
but hadronic degrees of freedom, are more natural. 

As said above, this matching takes into account 
exactly all the short-distance scale and 
 scheme dependences as well as the choice of evanescent operators.
We are left with the coupling of the $X$-boson long-distance
effective action completely fixed in terms of
the SM ones
\ba
\frac{g_{\Delta S=2}^2(\mu_C,\cdots)}{M_X^2}
\equiv C_{\Delta S=2} \, C(\nu) \, \left[ 1 + a \left( \gamma_1 
\log\left(\frac{M_X}{\nu}\right) + \Delta r \right) \right] \, .
\ea
 The one-loop finite term  $\Delta r$ is scheme dependent
\be
\Delta r^{NDR} = -\frac{11}{8} \left(1-\frac{1}{N_c}\right) \, ; 
\Delta r^{HV} = -\frac{7}{8} \left(1-\frac{1}{N_c}\right) \,  
\ee
and makes the coupling $|g_{\Delta S=2}|$ scheme independent
to order $a^2$.  This coupling is as well  analytically
scale-$\nu$ independent  at the same order.
 Notice  that there is no dependence on 
the cut-off scale $\mu_C$ --this feature
is general of four-point functions  which are
product of conserved currents \cite{BP00}.

 In the procedure described above, one also produces the standard
 leading and next-to-leading resummation to all orders of 
the large logs in $[\alpha_S \log(M_W/\nu)]^n$ and 
$\alpha_S \, [\alpha_S \log(M_W/\nu)]^n$. And this has been done
in the two-schemes described before; namely, NDR and HV.

Once the long-distance effective action
${\bf \Gamma}_{LD}$ is fully fixed,
we are ready to calculate the relevant matrix element.
\ba
\langle {\overline K}^0(q) | e^{i {\bf \Gamma}_{\Delta S=2}}
| K^0 (q) \rangle &=&
\langle {\overline K}^0(q) | e^{i {\bf \Gamma}_{LD}}
| K^0 (q) \rangle \equiv -i C_{\Delta S=2} \, 
\frac{16}{3} \, \hat B_K \, q^2 f_K^2  \nn \\
&=& \int \frac{{\rm d}^4 p_X}{(2\pi)^4} \, 
\frac{g^2_{\Delta S=2}}{2} \, \frac{i g_{\mu\nu}}{p_X^2-M_X^2} \,
{\bf \Pi}^{\mu\nu}(p_X^2,q^2) \,   
\ea
where $q^2$ is the external momentum carried by the kaons.
The basic object is the four-point function
\be
\label{basic}
{\bf \Pi}^{\mu\nu}(p_X^2,q^2) \equiv i^2 \, 4 \,
\langle \overline{K}^0(q) | \int {\rm d}^4 x \, \int {\rm d}^4 y \, 
e^{-i p_X \cdot (x-y)} \, T\left( L^\mu(x) \, L^\nu(y)
\right) | K^0(q) \rangle
\, .
\ee
 At large $N_c$,  this four-point function factorizes
into two two-point functions at all orders in quark masses
and external momentum $q^2$.
The disconnected part of the four-point $\Delta S=2$ function is
\be
g_{\mu\nu} {\bf \Pi}^{\mu\nu}_{\rm disconn.}(p_X^2, q^2)
= (2\pi)^4 \delta^{(4)}(p_X) \, 8 \, q^2 \, f_K^2
\ee
which leads to the well-known large-$N_c$ prediction
\be
\hat B_K^{N_c}= \frac{3}{4} \, .
\ee

At next-to-leading in the $1/N_c$ expansion, one has
\be
\label{NLO}
\hat B_K = \frac{3}{4} \left[
1-\frac{1}{16\pi^2 f_K^2}
{\dis \int_0^\infty} {\rm d} Q^2 \, F[Q^2] \right]
\ee
with $Q^2$ the X-boson momentum in the Euclidean space and
\ba
\label{FF}
F[Q^2]&\equiv&
-\frac{g^2_{\Delta S=2}}{8 \pi^2} {\dis \lim_{q^2\to m_K^2}}
\int {\rm d} \Omega_Q \, \frac{Q^2}{Q^2+M_X^2}
\, \frac{
  g_{\mu\nu}{\bf \Pi}^{\mu\nu}_{\rm conn.} (Q^2, q^2)}{q^2} \, .
\ea

 The next point is the calculation of (\ref{FF}).
 There are two energy regimes where we know
how to calculate  $F(Q^2)$  within QCD. The first one,
it is when $Q^2 >> 1 \, {\rm GeV}^2$ while $q^2$ is kept
small since we will have to put it on the kaon mass-shell.
In this regime, using the operator product expansion
within QCD  one gets
\be
\label{OPEfour}
g_{\mu\nu} {\bf \Pi}^{\mu\nu}_{\rm conn.} (Q^2, q^2) = 
{\dis \sum_{n=2}^\infty} \frac{C^{(i)}_{2n+2}
(\nu,Q^2) 
\langle {\overline K}^0 (q) | {\bf Q}^{(i)}_{2n+2} |
{K}^0 (q) \rangle}{Q^{2n}}
\ee
where ${\bf Q}^{(i)}_{2n+2}$  are local $\Delta S=2$
operators of dimension $2n+2$. In  particular,
\be
{\bf Q}_6 = 4 \, \int {\rm d}^4 x \, L^\mu(x) L_\mu(x)
\ee
 and 
\ba
C_6(\nu,Q^2) = - 8 \pi^2 \gamma_1 a \,  
\left[1 + a \left[ \left(\beta_1-\gamma_1\right) \, 
\log \left( \frac{Q}{\nu} \right) + {\bf F}_1 \right] 
+ O(a^2) \right]
\ea
with
\be
{\bf F}_1= \frac{\gamma_2}{\gamma_1}
+\left( \beta_1 -\gamma_1 \right)
\, \left[\frac{\Delta r}{\gamma_1} - \frac{1}{2} \right] \, 
\ee
where the $a^2$ term was not known before.
The finite term ${\bf F}_1$ is order $N_c$  and therefore this
 $a^2$ term is of the same order  as the leading term. In fact, at the
same order in $N_c$ there is an infinite series in powers of $a$.

 For the list and a discussion of the dimension eight operators
see \cite{CP03,CDG00}. In \cite{CP03} there is a calculation
in the factorizable limit of the contribution of the dimension eight 
operators.  Numerically,   the  finite term of order $a^2$ competes
with that result when  $Q^2$ is around $(1\sim 2)$ GeV$^2$.

The second energy regime where  we can calculate $F[Q^2]$
model independently  is for $Q^2 \to 0$ where
the effective quantum field theory  of QCD is 
chiral perturbation theory. The result is known \cite{BP00,PR00} up to
order $p^4$ 
\be
\label{slope}
F^\chi[Q^2] = 3 - \frac{12}{F_0^2} \left(
2 L_1 + 5 L_2 + L_3 + L_9 \right) \, Q^2 + \cdots
\ee
with $F_0$ the chiral limit of the pion decay constant
$f_\pi=92.4$ MeV.

 We still need the intermediate energy region
for which we use the large $N_c$ hadronic  model described 
in the next section.

\subsection{Large $N_c$ Hadronic Model}

Up to this point, all the results are model independent.
In particular, we have seen in the previous section that there
are two energy regimes which can be calculated within QCD.
In this section, we describe a large $N_c$ hadronic model
which will provide the full ${\bf \Pi}^{\mu\nu}_{\rm conn.}(Q^2, q^2)$
which  contains these two regimes analytically.

\begin{figure}[ht]
\centerline{\epsfxsize=4.1in\epsfbox{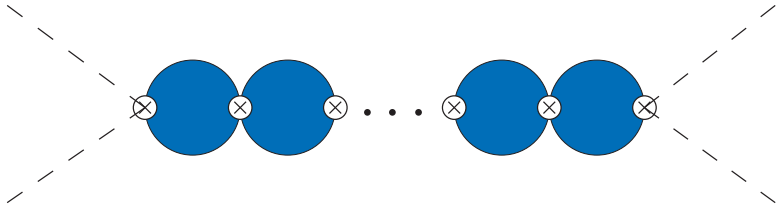}}   
\caption{ Infinite geometrical series which gives 
full two-point functions at large $N_c$. The crosses
that glue the vertex functions are either $g_V$
for vector or axial-vector sources or $g_S$ for scalar and
pseudo-scalar sources.
 \label{fulltwopoint}}
\end{figure}

 The large $N_c$ hadronic model we use was introduced in
\cite{BGLP03}.  The basic objects are vertex functions
with one, two, $\ldots$ two-quark
currents or density sources attached to them, referred to as one-point,
two-point,$\ldots$ vertex functions.
\begin{figure}[ht]
\centerline{\epsfxsize=4.1in\epsfbox{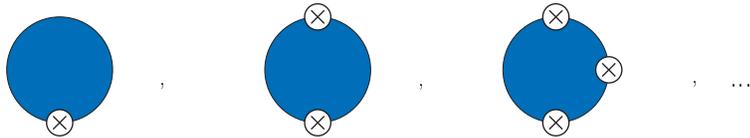}}   
\caption{ One-point, two-points, three-points, $\cdots$ vertex functions.
The crosses can be vector or axial-vector currents, scalar or pseudoscalar
densities.
 \label{vertex}}
\end{figure}
These vertex functions are glued into infinite
geometrical series with couplings $g_V$ for vector or
axial-vector sources and $g_S$ for scalar or pseudo-scalar
sources. In these way one can construct full n-point
Green's functions in the presence of quark masses
--see for instance, how to get full two-point functions
in Figure \ref{fulltwopoint}.

   The basic vertex functions in Figure \ref{vertex}
have to be polynomials in momenta and quark masses to keep 
the large $N_c$ structure but just keeping
the first octet of hadronic states  per channel, 
i.e., the pseudo-scalar pseudo-Goldstone bosons, the first vectors, 
the first axial-vectors and 
the first scalars non-dynamically generated resonances.
 The coefficients of the vertex functions are free constants
of order $N_c$. Imposing chiral Ward identities on the full
Green's functions one can  fix most of these unknowns. 
 Chiral perturbation theory at order $p^4$ and the operator
product expansion in QCD  help to fix the rest of the free coefficients
of the vertex functions.

The full two-point Green's functions obtained from the resummation
in Figure~\ref{fulltwopoint}
agree with the ones of large $N_c$ when one limits the hadronic
content to be just one hadronic state per channel 
--our model does not produce any less or more constraints 
and all parameters can be fixed
in terms of resonance masses \cite{BP00}. 
Introducing two or more hadronic states per channel systematically
could be done --it would just be much more cumbersome.
To our knowledge, 
all the low-energy hadronic effective actions used 
for large $N_c$ phenomenology are in the  approximation
of keeping the first resonances below some hadronic scale
and in some cases  not in all channels.

All the process of obtaining  full Green's functions
can be done in the presence of quark masses.
 In fact, in \cite{BP00}  two-point functions were calculated outside
the chiral limit  and all the new parameters  that appear up
 to order $m_q^2$ can be fixed except one;
namely,  the second derivative of the quark condensate with respect
to quark masses.  Some predictions of the model we are discussing
involving coupling constants and masses
of vectors and axial-vectors  in the presence of masses  are
\ba
f^2_{Vij} \, M^2_{Vij} &=& f^2_{Vkl} \, M^2_{Vkl} \, , \nn \\
f^2_{Vij} \, M^4_{Vij} - f^2_{Vkl} \, M^4_{Vkl} &=& 
-\frac{1}{2} \langle \overline q q \rangle_\chi \, 
\left(m_i+m_j-m_k-m_l\right)\, \nn \\
f^2_{Aij} \, M^2_{Aij} + f^2_{ij}&=& f^2_{Akl} \, M^2_{Akl} 
+ f^2_{kl}\, , \nn \\
f^2_{Aij} \, M^4_{Aij} - f^2_{Akl} \, M^4_{Akl} &=& 
\frac{1}{2} \langle \overline q q \rangle_\chi \, 
\left(m_i+m_j-m_k-m_l\right)\, , 
\ea
where $i,j,k,l$ are indices for the up, down
and strange quark flavors.

\begin{figure}[ht]
\centerline{\epsfxsize=4.1in\epsfbox{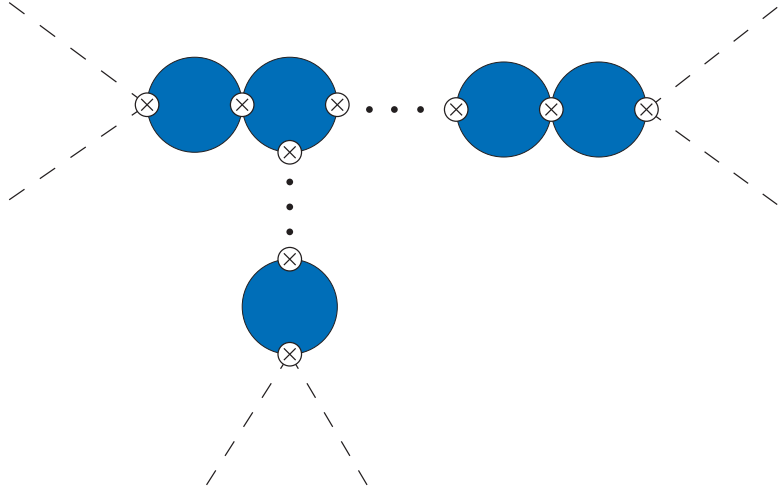}}   
\caption{ Infinite geometrical series which gives 
full three-point functions at large $N_c$. The crosses
that glue the vertex functions are either $g_V$
for vector or axial-vector sources or $g_S$ for scalar and
pseudo-scalar sources.
 \label{fullthreepoint}}
\end{figure}

Three-point functions  (see Figure \ref{fullthreepoint})
are just calculated at  present in the chiral limit
-- outside the chiral limit three-point functions 
will be ready soon \cite{BGP04}. 
Some three-point functions have also been
calculated  in other large $N_c$ approaches also in the chiral limit
\cite{MOU95,KN01,CIR04,RPP03}
--for instance, PVV, PVA, PPV and PSP three-point functions, where
P stands for pseudoscalar, V for vector, A for axial-vector
and S for scalar sources.
 They agree fully with the ones we get in our model
when restricted to just one hadronic state per channel.
As we said before one could also add systematically 
more hadronic states per channel in our large $N_c$ model.

 For the moment, we have just calculated 
the full four-point functions --and the corresponding
four-point vertex functions--  needed for (\ref{basic})
and in the chiral limit but we are finishing their calculation  outside 
the chiral limit too. 

\section{Results}

After integrating over the four-dimensional Euclidean
solid angle $\Omega_Q$ and doing the limit $q^2 \to 0$
as in (\ref{FF}), we get
\ba
\label{FFchi}
F^\chi[Q^2] &=& \frac{\alpha_V}{Q^2+M_V^2} + \frac{\alpha_A}{Q^2+M_A^2}
+ \frac{\alpha_S}{Q^2+M_S^2}
+ \frac{\beta_V}{\left(Q^2+M_V^2\right)^2}
 \nn \\ 
&+& \frac{\beta_A}{\left(Q^2+M_A^2\right)^2}
+ \frac{\gamma_V}{\left(Q^2+M_V^2\right)^3} +
\frac{\gamma_A}{\left(Q^2+M_A^2\right)^3} \, .
\ea
This is the input needed in (\ref{NLO}) to get $\hat B_K^\chi$.
The explicit calculation  reveals  that diagrams with the exchange 
of vector and axial-vector states produce not only single poles
as one may naively assume but two- and three-poles terms
 in those states masses too. On the opposite, scalar
exchange just produce single poles --this is due to the different powers of
 momenta involved in the scalar vertices with respect to the 
spin one ones.

 The function $F^\chi[Q^2]$ in (\ref{FF})
reproduces the large $N_c$-pole structure
 found in \cite{PR00,CP03} but including 
the first hadronic state in all the channels.
 The coefficients $\alpha_A$, $\alpha_V$, $\alpha_S$,
$\beta_V$, $\beta_A$, $\gamma_V$, and  $\gamma_A$ are known
functions of $M_V^2$, $M_A^2$, $M_S^2$, $F_0^2$ and the four
unknowns we mentioned above.  We need now to fix these inputs.
  
For $M_V$, $M_A$ and $M_S$, one could  either identify them with 
 the measured  masses of the corresponding lowest lying hadronic states.
 That means the first multiplet for the vector and axial-vector multiplet
while in the scalar case, the second multiplet is favored in view
of the increasing evidence for a dynamical origin of the first
-see \cite{CENP03} for a recent discussion of this issue.
Another possibility is to use the one to one large $N_c$ relation
to write the resonance masses in terms of the order $p^4$ CHPT 
couplings $L_i$.  In the large $N_c$ limit both choices are identical.
We prefer to take the second strategy since we want  
to have the order $p^4$ term in the CHPT expansion exact
analytically. We use $L_9=5.9 \cdot 10^{-3}$,
$L_{10}=-4.7\cdot 10^{-3}$, $L_5=0.9\cdot 10^{-3}$ from
 \cite{ABT01,BT02} which correspond to $M_V=0.79$ GeV, $M_A=1.03$ GeV,
and $M_S=1.43$ GeV, which agree with the experimental
masses  within the typical  $1/N_c$ uncertainty around 30 \%.

\begin{figure}[t]
\centerline{\epsfxsize=2.2in
\rotatebox{270}{\epsfbox{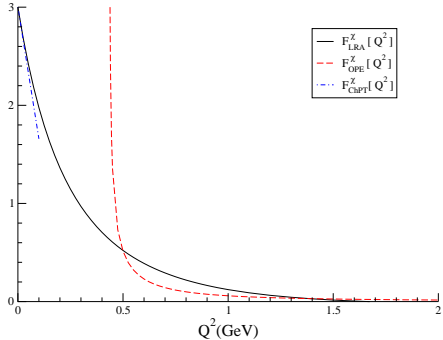}}}  
\caption{We plot the function $F^\chi[Q^2]$
in three cases: the full case in (\ref{FFchi})
is the solid line, the OPE result in (\ref{OPEfour}) 
is the dashed line and the CHPT result in (\ref{slope}) 
is the dot-dashed line. 
 \label{FFplot}}
\end{figure}
 The four unknowns we are left with correspond to three-point functions, 
namely one for the PSS, one for the PAS and the other two
for the PVA three-point function.  These we fix  as follows:
we fix the one related to the PSS three-point function  requiring 
that the slope at the origin is  the one in (\ref{slope}); the
constant related to the PAS three-point function is fixed by 
requiring that
 the residue of the $1/Q^2$ term in the OPE in (\ref{OPEfour})
vanishes. The two constants related to the three-point function
PVA are set by  requiring  at the same time  good matching between
our $F^\chi[Q^2]$ and the OPE result including dimension six and eight 
\cite{CP03} and  that this matching is between 1 and 2 GeV$^2$.
We checked that 
the dependence on the exact point of the matching is negligible.
That fixes completely all the unknowns.

Notice that  we have analytical scale and scheme independence
as was explained in Section \ref{Technique}.
 The result for $F^\chi[Q^2]$ is plotted in Figure \ref{FFplot}.

\begin{figure}[t]
\centerline{\epsfxsize=1.8in
\rotatebox{270}{\epsfbox{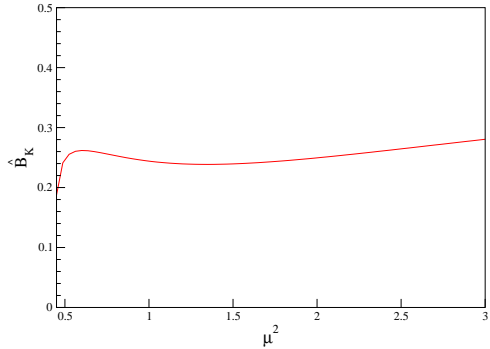}}}   
\caption{$\hat B_K^\chi$ plotted vs the upper
limit of the integral in (\ref{FF}). See text for 
further explanation.
 \label{BKplot}}
\end{figure}

 We now use (\ref{NLO})  and integrate 
(\ref{FF}) up to the matching point $\mu$ and from that
point on we use the OPE result including dimension
eight corrections \cite{CP03}. This OPE contribution to 
$\hat B_K^\chi$ is
negligible. In Figure \ref{BKplot}, we plot 
the value one obtains for $\hat B_K^\chi$ integrating 
(\ref{FF})  up to the matching scale $\mu$.
Notice the nice plateau one gets between 1  and 2 GeV$^2$.
Varying all the inputs: values of $L_i$s needed, $\alpha_S$,
$\cdots$ between their respective uncertainties we get as preliminary
result
\be
\hat B_K^\chi = 0.29 \pm 0.15 \, 
\ee
which is somewhat smaller  but fully compatible with the one found
previously also in the chiral limit in \cite{BP95,BP00,PR00,CP03}.

\section{Prospects and Conclusions}

The full analysis of $\hat B_K$ including chiral corrections  is
 under way. The results of the two energy regimes explained in 
Section \ref{Technique} 
are known. It is easy to get that $F[0]=0$ for the real case 
instead  of $F^\chi[0]=3$ in the chiral limit. Notice 
that the value of $F^\chi[0]$ is a strong constraint on  the value
of $\hat B_K^\chi$.
The CHPT calculation in the real quark masses case
$F[Q^2]$ has been done to order $p^4$ \cite{BP03}
and  $F[Q^2]$ remains small  --below 0.15-- up to energies
around  0.2 GeV$^2$ and then goes negative. 
The other energy regime in which $F[Q^2]$
is also known is for very large values of $Q^2$,  
which can be calculated using the OPE in QCD.
 In fact, the dimension six operator and Wilson coefficient
are the same as the chiral limit one.

One expects that higher CHPT order terms correct the behaviour
of $F[Q^2]$ for $Q^2>$ 0.2 GeV$^2$
and $F[Q^2]$ will  tend to the chiral limit curve
in Figure \ref{FFplot}. The  question  is at which energy does it happens. 
This can only be answered using a hadronic large $N_c$
approximation to QCD at present. So, though we have strong indications
that the value $\hat B_K=3/4$ has small chiral corrections as
shown before, we have to wait till we get the full four-point
 Green's  function in the real case to confirm it(\ref{basic}). 

 In the past few years,  lattice QCD  has produced many calculations
using  different fermion formulations, including preliminary unquenched
studies --see \cite{lattice04a,lattice04b} for references. 
The recommended lattice world average result in \cite{lattice04b} is  
$\hat B_K=0.81^{+0.06}_{-0.13}$.  There have also appeared chiral limit  
extrapolations --like for instance  in \cite{staggered} 
obtaining $\hat B_K^\chi=0.32\pm0.22$ or in \cite{AOK04}
obtaining $\hat B_K^\chi=0.34\pm0.02$
--which show a clear decreasing tendency 
with respect to the real case value, in agreement with the 
results found here and in \cite{BP95,BP00,PR00,CP03}.

The  QCD-Hadronic Duality result for $\hat B_K$\cite{PR85,PRA91} 
 is very close to the chiral limit result above because 
--as already mentioned in \cite{PRA91}-- what
was calculated there, is the order $p^2$ coefficient
of the chiral expansion which actually is the chiral limit value
of $\hat B_K$.

\section*{Acknowledgments}

J.P. thanks the Department of Theoretical
Physics at Lund University where part of his work was done 
for the warm hospitality.
 This work has been supported in part by
the European Comission (EC) RTN Network EURIDICE
Grant No.  HPRN-CT-2002-00311 (J.B. and J.P.),
by Swedish  Science Foundation,
by EC Marie Curie Grant No. MEIF-CT-2003-501309
(E.G.), by MCYT (Spain) and FEDER (EC) Grant No.
FPA2003-09298-C02-01 (J.P.), and by the Junta
de Andaluc\'{\i}a Grant No. FQM-101 (J.P.).

\end{document}